\documentclass[twocolumn,showpacs,floats,floatfix,superscriptaddress,aps,pra]{revtex4}
\usepackage{amsfonts}
\usepackage{amssymb}
\usepackage{amsmath}
\usepackage{graphicx}
\usepackage{bm}
\usepackage{color}
\usepackage{epsfig}
\usepackage{ifthen}
\usepackage{color}
\usepackage{hyperref}

\def\bfit#1{\textbf{\textit{#1}}}
\def\bfit#1{\subsubsection{#1}}
\def\e{\,\text{e}}
\def\d{\,\text{d}}
\def\i{\,\text{i}}
\def\sech{\,\text{sech}\,}
\def\mycite{\cite}

\def\be{\begin{equation}}
\def\ee{\end{equation}}
\def\bea{\begin{eqnarray}}
\def\eea{\end{eqnarray}}
\def\bse{\begin{subequations}}
\def\ese{\end{subequations}}

\def\calE{\mathcal{E}}

\def\bmat{\left[ \begin{array}{ccc}}
\def\emat{\end{array} \right] }

\def\A{\mathcal{A}}

\def\H{\mathsf{H}}

\def\Re{\,\text{Re}\,}
\def\Im{\,\text{Im}\,}
\def\sech{\,\text{sech}\,}
\def\bc{\begin{center}}
\def\ec{\end{center}}

\begin{document}

\title{Rapid adiabatic passage without level crossing}

\author{A. A. Rangelov}
\affiliation{Department of Physics, Sofia University, James Bourchier 5 blvd., 1164 Sofia, Bulgaria}
\author{N. V. Vitanov}
\affiliation{Department of Physics, Sofia University, James Bourchier 5 blvd., 1164 Sofia, Bulgaria %
 and Institute of Solid State Physics, Bulgarian Academy of Sciences, Tsarigradsko chauss\'{e}e 72, 1784 Sofia, Bulgaria}
\author{B. W. Shore}
\affiliation{618 Escondido Cir., Livermore, CA 94550, USA} %

\date{\today}
\begin{abstract}
We present a method for achieving complete population transfer in
a two-state quantum system via adiabatic time evolution in which,
contrary to conventional rapid adiabatic passage produced by
chirped pulses, there occurs no crossing of diabatic energy
curves: there is no sign change of the detuning. Instead, we use
structured pulses, in which, in addition to satisfying conditions
for adiabatic evolution, there occurs a sign change of the Rabi
frequency when the detuning is zero. We present simulations that
offer simple geometrical interpretation of the two-dimensional
motion of the Bloch vector for this system, illustrating how both
complete population inversion and complete population return occur
for different choices of structured pulses.
\end{abstract}


\maketitle

\section{Introduction \label{Sec-Intro}}

The technique of rapid adiabatic passage (RAP) \cite{All87,Sho90,Vit01a,Sho08,Vit01b} provides a well-studied robust and
efficient method for producing complete population transfer between two bound states of a quantum system. Traditionally the
population change has been accomplished by sweeping the carrier frequency of a laser pulse through resonance with a two-state
system (parameterized by detuning), while simultaneously pulsing the strength of the interaction (parameterized as time-varying
Rabi frequency). The sweep of detuning corresponds to a crossing of diabatic energy curves, as noted below. When the interaction
changes sufficiently slowly, so that the time evolution is adiabatic, there will occur a complete population transfer from the
ground state (of energy $E_1$) to the excited state (energy $E_2$). %
Alternatively, one can make the detuning cross resonance by sweeping the transition frequency, e.g. by time-dependent electric
or magnetic fields, as exemplified by the recently introduced technique of Stark-chirped rapid adiabatic passage (SCRAP)
\cite{SCRAP}.

In this paper we exploit a novel way of achieving complete
population transfer via adiabatic evolution in two-state systems,
using pulses that produce no crossing of diabatic energies. Such
population transfer is possible when diabatic energy curves touch,
but do not cross, and simultaneously there occurs a sign change of
the interaction (the Rabi frequency).

\section{Background \label{Sec-RAP}}

Coherent excitation of a two-state quantum system is described by
the time-dependent Schr\"{o}dinger equation (TDSE), which in the
rotating-wave approximation (RWA) \cite{All87,Sho90} reads
\begin{equation}
i\hbar\frac{d}{dt}\mathbf{C}(t) = \H(t)\mathbf{C}(t),
\label{SEq-c}
\end{equation}
where $\mathbf{C}(t) = [C_1(t),C_2(t)]^T$ is a two-dimensional column-vector whose elements are the probability amplitudes
$C_1(t)$ and $C_2(t)$ associated with the two quantum states $\psi_1$ and $\psi_2$, and $\H(t)$
 is the matrix representing the RWA Hamiltonian \cite{All87,Sho90},
\begin{equation}\label{Hc}
\H(t) = \frac \hbar 2 \left[ \begin{array}{cc} -\Delta (t) & \Omega (t) \\ \Omega (t) & \Delta (t)
\end{array}\right].
\end{equation}
The detuning $\Delta(t) = \omega_0(t)-\omega_L(t)$ expresses the frequency offset of the laser carrier frequency $\omega_L(t)$
from the Bohr transition frequency $\omega_0(t)$, each of which may vary with time. The Rabi frequency $\Omega(t)$ quantifies
the field-induced coupling between the two states. For the laser-atom excitation considered here this is the product of the
atomic dipole transition moment $d_{12}$ in the field polarization direction, and the electric-field envelope $\calE(t)$ of the
laser field, $\Omega (t)= -d_{12}\calE(t)/\hbar$. We shall take this to be a real-valued function of time.

In writing the RWA Hamiltonian we have chosen a symmetrized form, which is obtained by a suitable phase transformation of the
probability amplitudes. The diagonal elements of $\H (t)$ are \emph{diabatic energies}. Curves of diabatic energies cross when
$\Delta(t)$ changes sign. The (time-dependent) eigenvalues of $\H (t)$ are \emph{adiabatic energies},
$\hbar\varepsilon_{\pm}(t)$. By definition, $\varepsilon_+(t)$ is the larger of the
two eigenvalues. 

As has long been known \cite{Feynman}, Eq \eqref{SEq-c} for the two
complex-valued probability amplitudes can be recast as three coupled
equations for real-valued variables, that serve as the three
components of a unit vector $\mathsf{B}(t)=[u(t),v(t),w(t)]^T$ in a
three-dimensional abstract space. The components of this \emph{Bloch
vector} are
\bse\bea
 && u(t) = 2\Re \rho_{12}(t),\\
 && v(t) = 2\Im \rho_{12}(t),\\
 && w(t) = \rho_{22}(t) - \rho_{11}(t),%
\eea\ese%
with $\rho_{mn}(t) = C_m^\ast(t) C_n(t)$ ($n,m=1,2$) being the matrix elements of the density operator. The resulting RWA
optical Bloch equation reads \cite{All87,Sho90,Feynman}
\begin{equation} \label{Bloch equation}
\frac{\,\text{d}}{\,\text{d} t} \mathbf{B}(t) %
= \mathsf{R}(t) \mathbf{B}(t),
\end{equation}
with
\begin{equation} \label{R}
\mathsf{R}(t) = \left[\begin{array}{ccc}
0 & -\Delta (t) & 0 \\
\Delta (t) & 0 & -\Omega (t) \\
0 & \Omega (t) & 0
\end{array}\right].
\end{equation}
The skew-symmetric nature of the matrix $\mathsf{R}(t)$ allows us
to write Eq ~\eqref{Bloch equation} as a torque equation for the
Bloch vector \cite{Feynman,Ran09},
\begin{equation}\label{eqn-torque}
\frac{\,\text{d}}{\,\text{d} t}\mathbf{B}(t)=\mathbf{Q}(t)\times\mathbf{ B}(t),
\end{equation}
where the driving torque (angular velocity) vector $\mathbf{Q}(t)$ reads \cite{Feynman,Ran09}
\begin{equation}\label{torque vector}
\mathbf{Q}(t) = [\Omega(t),0,\Delta(t)]^T.
\end{equation}
The Bloch vector rotates at the instantaneous rate%
\be%
 \widetilde{\Omega}(t) = \sqrt{\Omega(t)^2 + \Delta(t)^2}.
\ee%
Note that the direction of rotation will be reversed if the signs of $\Omega(t)$ and $\Delta(t)$ are both reversed.

One of the three eigenvectors of $\mathsf{R}(t)$, which corresponds
to a null eigenvalue, is a sum of only two components, the coherence
$u(t)$ and the inversion $w(t)$,
\begin{equation}\label{pseudo-dark state}
 a_0(t) = u(t)\sin\vartheta(t) + w(t)\cos\vartheta(t).
\end{equation}
It has no component of $v(t)$; in analogy with stimulated Raman adiabatic passage (STIRAP) \cite{Gau90,Ber98,Vit06}, we shall
refer to it as a \emph{dark vector}. %
Here the time-dependent mixing angle $\vartheta(t)$ is defined by
\begin{equation}
\tan\vartheta (t) = \Omega(t)/\Delta(t).
\end{equation}
In contrast to the dark state of STIRAP, the dark vector $a_0(t)$ can radiate via spontaneous emission from the excited state
$\psi_2$. It is important to note that the components of the dark vector are the normalized components of the torque vector
\eqref{torque vector}.

Below we describe several situations in which the time evolution is adiabatic, so that the Bloch vector \emph{adiabatically
follows} the torque vector as, in accord with the design of the functions $\Delta(t)$ and $\Omega(t)$, it moves in a
two-dimensional plane of the abstract vector space. The pulses can be designed to produce either complete population inversion
(CPI), in which all population resides in state 2 at the completion of the pulses, or complete population return (CPR), in which
the initially populated state is completely repopulated after the pulses.

\section{Adiabatic following with level crossing \label{Sec-crossing}}

\begin{figure}[tb]
\epsfig{width=80mm,file=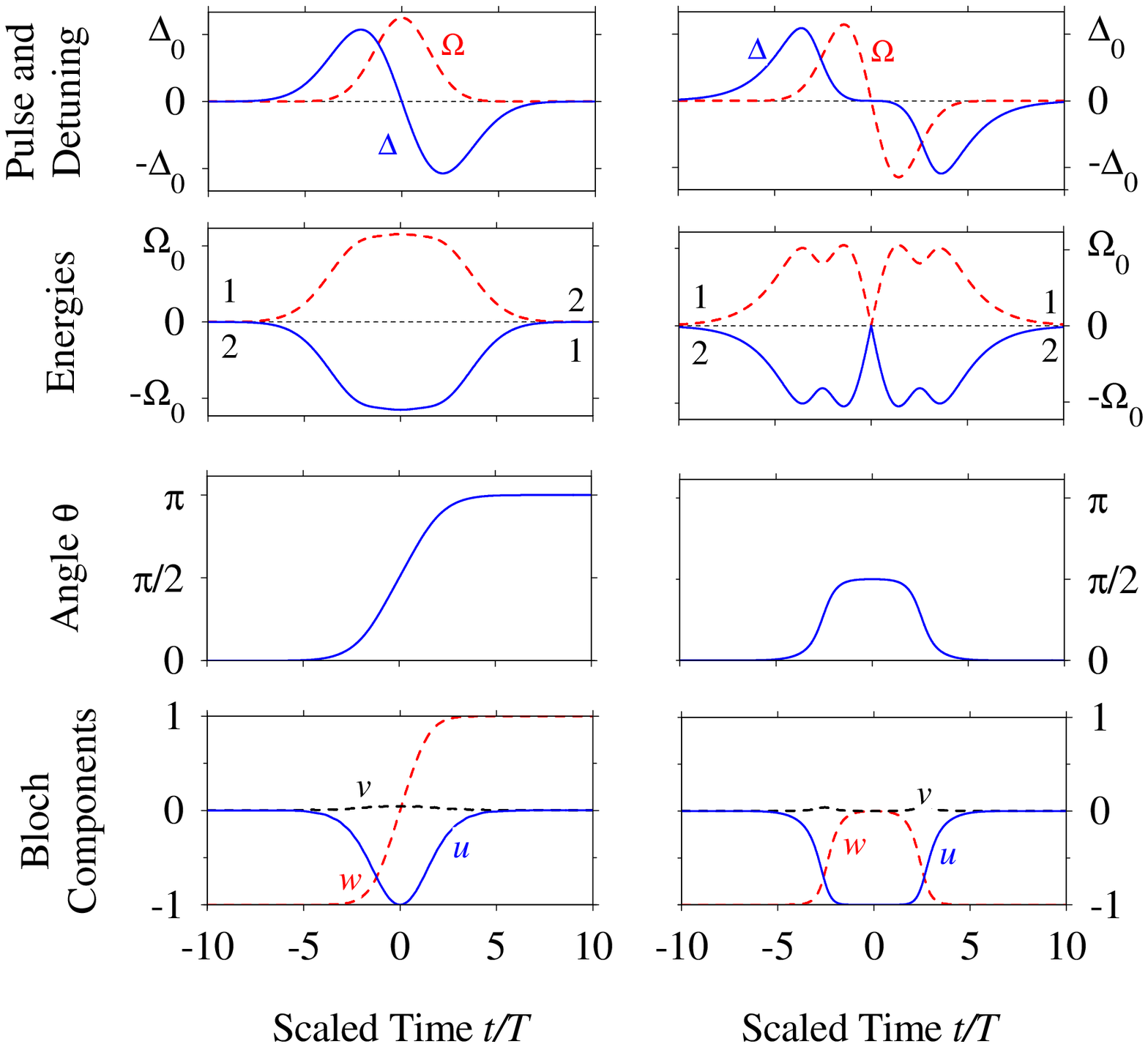} \caption{CPI (left frames) and
CPR (right frames) in the
presence of level crossing (detuning crossing resonance). \emph{Left frames}%
: The detuning and the Rabi frequency are given by Eqs
~\eqref{frame a}, with $\Delta _{0}=-40/T$ and $\Omega _{0}=30/T$.
\emph{Right frames}: the
detuning and the Rabi frequency are given by Eqs ~\eqref{frame b}, with $%
\protect\tau =3T$, $\Delta _{0}=-25/T$ and $\Omega _{0}=-80/T$.
\emph{Top frames}: the Rabi frequency and the detuning;
\emph{second frames}: eigenenergies; \emph{third frames}: mixing
angle $\vartheta (t)$; \emph{bottom frames}: components of the
Bloch vector $\mathbf{B}(t)$. } \label{FIG-RAP1}
\end{figure}

Figures \ref{FIG-RAP1} shows examples of adiabatic CPI or CPR in the presence of a level crossing (the detuning crosses the
resonance). The frames in the left column of Fig.~\ref{FIG-RAP1} illustrate CPI for the conventional rapid adiabatic passage
(RAP) by a frequency chirped laser pulse \cite{All87,Sho90,Vit01a,Vit01b,Sho08}. %
Top frames show the pulsed Rabi frequency and detuning as a function of time, for which we have chosen %
\bse\label{frame a}\bea%
&&\Delta(t) = \frac {\Delta_0 t}{T} \e^{-(t/1.5T)^2},\\
&&\Omega(t) = \Omega_0 \e^{-(t/T)^2}, %
\eea\ese%
The second frames show the eigenenergies, the third frames show the mixing angle $\vartheta(t)$,
 and the bottom frames show the components of the Bloch vector. %

The frames in the right column of Fig.~\ref{FIG-RAP1} illustrate the situation when, in addition to the level crossing of the
detuning, the Rabi frequency also changes sign,
\bse\label{frame b}\bea%
&&\Delta(t) = \frac {\Delta_0 t^2}{T^2} \left[\sech\left(\frac{t-\tau}T\right) - \sech\left(\frac{t+\tau}T\right)\right],\\
&&\Omega(t) = \frac {\Omega_0 t}{T} \e^{-(t/T)^2},
\eea\ese%
The temporal pulse area -- the time integral of the Rabi frequency -- is here zero (a ``zero-area pulse").
For this choice of antisymmetric Rabi frequency and detuning, exact
CPR takes place. Indeed, this property, which stems from the overall
antisymmetry of the Hamiltonian, is valid not only in the adiabatic
limit but in the general case, and for arbitrarily many states
\cite{Vitanov-Knight}.

For the left frames, during the excitation the mixing angle $\vartheta(t)$ rotates from $\vartheta(-\infty)=0$ to
$\vartheta(0)=\pi/2$ and then to $\vartheta(\infty)=\pi$ (due to the change of the sign of the detuning), which leads to CPI
because the system remains in the dark vector \eqref{pseudo-dark state}, which follows adiabatically the driving torque. For the
right frames, during the excitation the mixing angle $\vartheta(t)$ rotates from $\vartheta(-\infty)=0$ to $\vartheta(0)=\pi/2$
and then back to $\vartheta(\infty)=0$ (due to the change of sign of the detuning and the Rabi frequency at the crossings),
which leads to CPR because, again, the system remains in the dark vector \eqref{pseudo-dark state}, but now it has different
asymptotics.

The choice of peak amplitudes for the pulses in these examples is such that a measurable value of Bloch vector component $v(t)$
occurs during intermediate stages of the process. As the motion becomes more adiabatic, and the adiabatic following more
complete, this component becomes smaller.

\section{Adiabatic following without level crossing \label{Sec-no-crossing}}

\begin{figure}[tb]
\epsfig{width=80mm,file=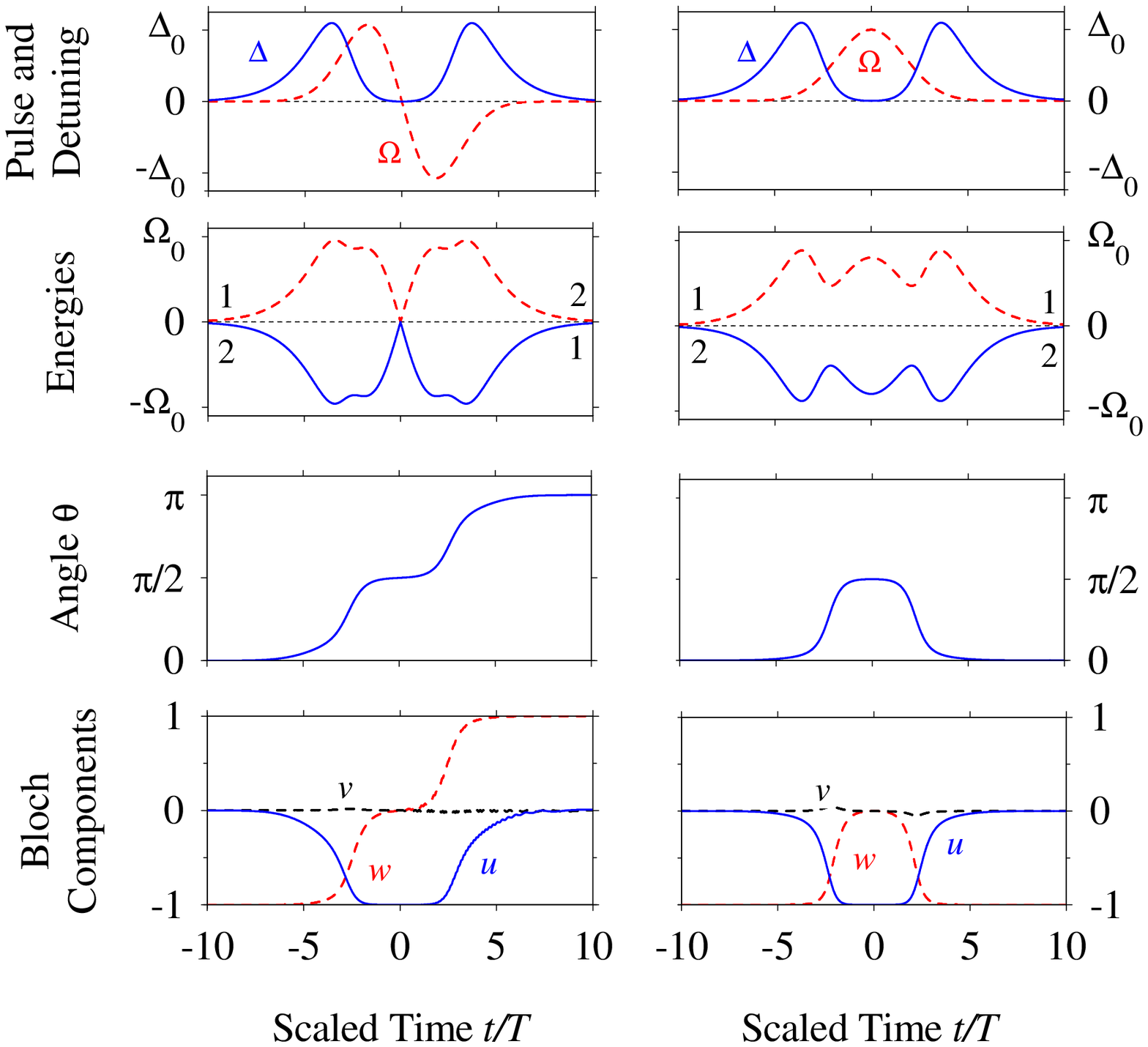} \caption{CPI (left frames) and
CPR (right frames) in the absence of level crossing (detuning only
touching resonance). \emph{Left frames}: the detuning and the Rabi
frequency are given by Eqs~ \eqref{frame c}, with $\protect\tau
=3T$, $\Delta _{0}=4/T$ and $\Omega _{0}=-40/T$. \emph{Right
frames}: the detuning and the Rabi frequency are given by
Eqs~\eqref{frame d}, with $\protect\tau =3T$, $\Delta _{0}=4/T$
and $\Omega _{0}=40/T$. \emph{Top frames}: the Rabi frequency and
the detuning;
\emph{second frames}: eigenenergies; \emph{third frames}: mixing angle $%
\protect\vartheta (t)$; \emph{bottom frames}: components of the
Bloch vector $\mathbf{B}(t)$.} \label{FIG-RAP2}
\end{figure}

Figure \ref{FIG-RAP2} shows a pair of cases, which do not appear to have been reported previously. In the left frames of
Fig.~\ref{FIG-RAP2} we illustrate CPI in the absence of a level crossing, when the detuning does not cross but just touches the
resonance. The detuning and the Rabi frequency read
\bse\label{frame c}\bea%
&&\Delta(t) = \frac {\Delta_0 t^2}{T^2} \left[\sech\left(\frac{t+\tau}T\right) + \sech\left(\frac{t-\tau}T\right)\right],\\
&&\Omega(t) = \frac {\Omega_0 t}{T} \e^{-(t/2.5T)^2},
\eea\ese%
During the excitation the mixing angle $\vartheta(t)$ rotates from
$\vartheta(-\infty)=0$ to $\vartheta (\infty) = \pi$, which leads
to CPI in the adiabatic limit because the system remains in the
dark vector \eqref{pseudo-dark state}. Furthermore for the given
example of antisymmetric pulse shape, the time-integrated Rabi
frequency (the temporal pulse area) is zero, but the zero pulse
area is not a necessary constraint. This is a new approach to CPI,
without a level crossing and with a possible zero pulse area.

The right frames of Fig.~\ref{FIG-RAP2} illustrate the case when the
Rabi frequency and the detuning are both even functions of time,
\bse\label{frame d}\bea%
&&\Delta(t) = \frac {\Delta_0 t^2}{T^2} \left[\sech\left(\frac{t+\tau}T\right)+\sech\left(\frac{t-\tau}T\right)\right],\\
&&\Omega(t) = \Omega_0 \e^{-(t/2.5T)^2},
\eea\ese%
The mixing angle $\vartheta(t)$ rotates from $\vartheta(-\infty)=0$ to $\vartheta(0)=\pi/2$ and then to $\vartheta(\infty)=0$,
which leads to CPI in the adiabatic limit because the system remains in the dark vector \eqref{pseudo-dark state}.

\section{Discussion \label{Sec-discussion}}

\subsection{General features \label{Sec-generality}}

In summary, the analysis above brings us to the following conclusions.
CPI takes place in the adiabatic limit when (i) the coupling is an even function of time and the detuning is odd
function of time ($\Omega(t)=\Omega(-t)$, $\Delta(t)=-\Delta(-t)$,
left frames of Fig.~\ref{FIG-RAP1}), or (ii) the coupling is odd
function of time and the detuning is
even function of time ($\Omega(t)=-\Omega(-t)$, $\Delta(t)=\Delta(-t)$), left frames of Fig.~\ref{FIG-RAP2}. %
When (iii) both functions are even function of time or (iv) both are odd function of time, then CPR is observed instead of CPI,
 right frames of Figure \ref{FIG-RAP1} and \ref{FIG-RAP2}. These findings are summarized in Table \ref{table}.

\begin{table}[tb]
\begin{center}
\begin{tabular}{|l|c|c|c|}\colrule
 Figure & Rabi frequency & detuning & adiabatic \\
 & $\Omega(-t)$ & $\Delta(-t)$ & limit \\ \colrule
 \ref{FIG-RAP1} (left) & $\Omega(t)$ & $-\Delta(t)$ & CPI \\
 \ref{FIG-RAP1} (right) & $-\Omega(t)$ & $-\Delta(t)$ & CPR \\
 \ref{FIG-RAP2} (left) & $-\Omega(t)$ & $\Delta(t)$ & CPI \\
 \ref{FIG-RAP2} (right) & $\Omega(t)$ & $\Delta(t)$ & CPR \\
\colrule
\end{tabular}
\caption{Adiabatic limit for fields of particular symmetry: CPI: complete population inversion; CPR: complete population
return.} \label{table}
\end{center}
\end{table}

We emphasize here that the assumption of definite symmetry of the Rabi frequency $\Omega(t)$ and the detuning $\Delta(t)$, which
was made in figures~\ref{FIG-RAP1} and \ref{FIG-RAP2}, and summarized in Table \ref{table}, is only a matter of convenience. In
the adiabatic following approximation, which was assumed throughout, it is only the \emph{asymptotic} behavior of the
\emph{mixing angle} $\vartheta(t)$, which controls the composition of the dark vector \eqref{pseudo-dark state}, that matters.

Because the adiabatic passage is robust, the procedure of CPI with level touching (left frames of Figure \ref{FIG-RAP2}) is also
robust: it depends only weakly on the overlap of the coupling $\Omega (t) $ and the detuning $\Delta (t)$ and their peak values.
Moreover, the exact antisymmetry is not required, and high efficiencies can also be achieved with asymmetric time dependences.
The only restriction of the pattern that we have to follow in order to achieve CPI is: the detuning starts before the coupling,
becomes zero before the coupling changes sign, then rises again with the same sign and ends after the coupling ends.

\begin{figure}[tb]
\epsfig{width=80mm,file=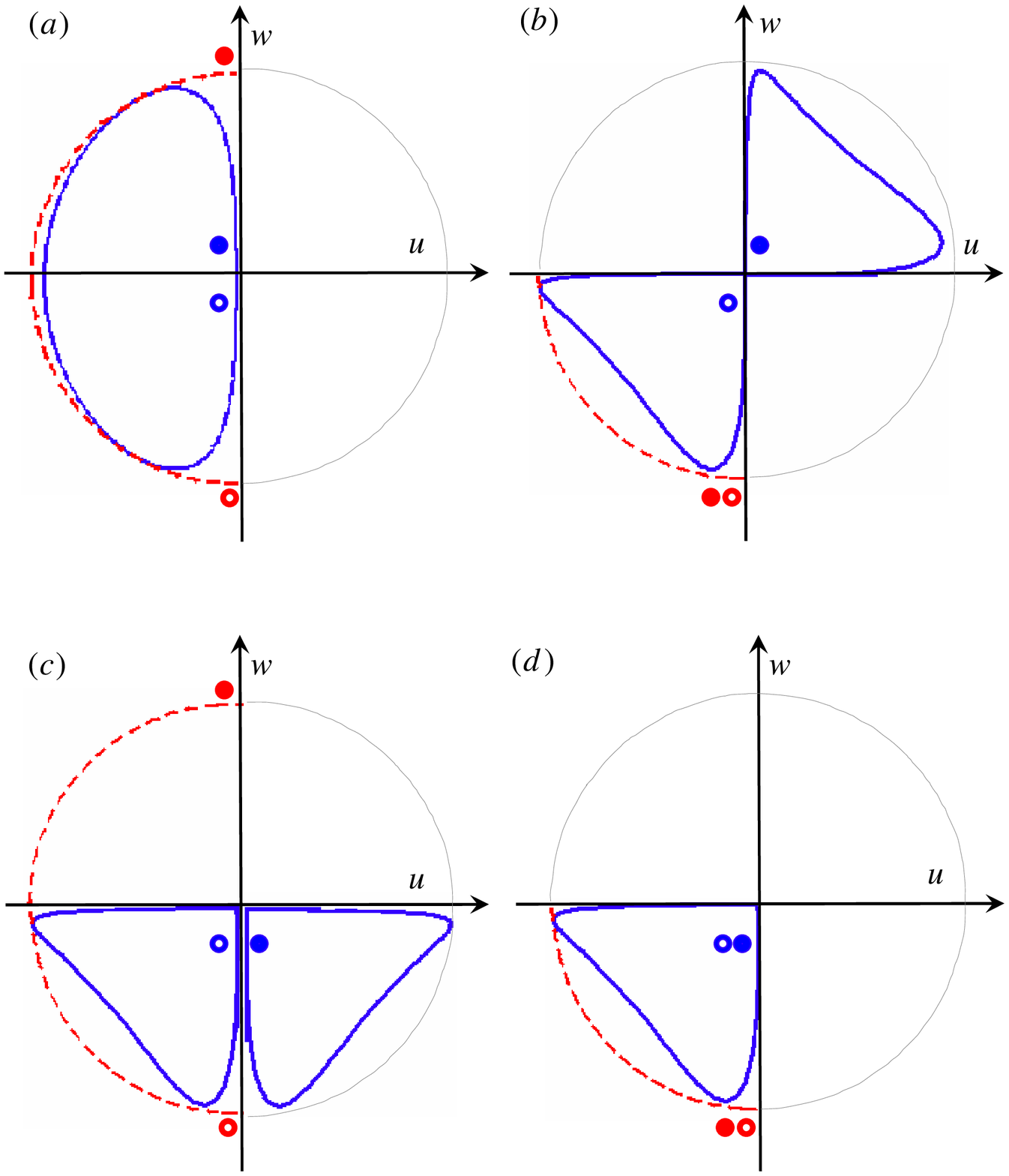} \caption{Motion of the Bloch
vector and the angular velocity vector in the $(u,w)$ plane. Solid
line (blue online) is the angular velocity (torque) vector,
normalized to peak value. Dashed line (red online) is the Bloch
vector. The initial positions of the Bloch vector and the torque
are marked by hollow circles. The final positions are marked by
filled circles. The cases (a)-(d) correspond to the cases of
figures \ref{FIG-RAP1} and \ref{FIG-RAP2}.} \label{FIG-Bloch
sphere}
\end{figure}

The torque-equation interpretation of the motion of the Bloch vector
provides a useful means of understanding the behaviors discussed
above.
Figure \ref{FIG-Bloch sphere} provides such a picture for the four frames of figures \ref{FIG-RAP1} and \ref{FIG-RAP2} and Table \ref{table}. %
In the adiabatic limit, which we assume, the Bloch vector evolves in the plane $(u,w)$, which is the $v=0$ slice of the Bloch sphere. %
The geometrical representation of the adiabatic evolution of the
Bloch vector is easily determined because in the adiabatic limit the
state vector coincides with the dark vector \eqref{pseudo-dark
state} and is always parallel or anti-parallel to the driving
torque.

\subsection{Experimental feasibility\label{Sec-feasibility}}

The concept of a zero-area laser pulse, when first encountered, may seem somewhat confusing. We have discussed such pulses in some
detail elsewhere \cite{Sho09}. Here we describe briefly a few of the available techniques.

\bfit{Nonoscillating magnetic fields.} Historically, the concept arose first in nuclear magnetic resonance (NMR), wherein
typically spin-$\frac12$ particles are controlled by slowly varying magnetic fields \cite{NMR}. A zero area field can be
produced in an obvious manner, by appropriate tailoring of an applied voltage that can reverse the direction of the field and
create the desired asymmetric pulse. In NMR the detuning $\Delta(t)$ is replaced by a longitudinal (along the quantization axis)
magnetic pulse, and the Rabi frequency $\Omega(t)$ by a transverse magnetic pulse.

\bfit{Self-induced transparency.} In the optical spectrum zero-area pulses are formed when pulses are reshaped by self-induced
transparency (SIT) while propagating through an absorbing medium. Initial pulse areas in the range $(0,2\pi)$ will tend toward
zero-area pulses -- an expression of the McCall-Hahn area theorem \cite{SIT}.

\bfit{Beam splitting \& recombination.} A zero-area pulse can also be produced by passing a laser pulse through a beam splitter
and then recombining the two split pulses, after delaying one of them by a time delay $\tau$ and shifting its phase by $\pi$.
The recombined pulse, with Rabi frequency
\begin{equation}
 \Omega (t) = \Omega_0 \left[ f(t-\tau)-f(t)\right] , \label{recombined}
\end{equation}
where $\tau$ is the delay, is a zero-area pulse. Such an implementation requires an interferometric stability within a
wavelength $\lambda$.

\bfit{Ultrashort shaped pulses.} %
The most useful techniques, for ultrashort (femtosecond and picosecond) laser pulses, manipulate the Fourier transform of the
pulse,
\begin{equation} \label{Fourier}
\widetilde{\Omega }(\Delta) = \frac{1}{2\pi} \int_{-\infty}^{\infty} \d t\,\e^{\i\Delta t}\,\Omega (t),
\end{equation}
using pulse shapers \mycite{femto}. The pulse area $\A$ is the value of this transform on resonance ($\Delta =0$),
\begin{equation}
A \equiv \int_{-\infty}^{\infty} \d t \, \Omega(t) = 2\pi
\widetilde{\Omega}(0). \label{area-Fourier}
\end{equation}
Thus a zero-area pulse can be produced by negating the resonant frequency component $\widetilde{\Omega}(0)$ in its spatially
dispersed Fourier spectrum.

The desired time variation of the detuning can be produced, using a
pulse shaper, \emph{simultaneously} with the manipulation of
the Rabi frequency. Alternatively, the pulse of Fig.~\ref{FIG-RAP2}(c) can be produced by shaping the Rabi frequency and the detuning \emph{separately}. 
For example, the Rabi frequency can still be shaped by a pulse shaper, whereas the detuning can be shaped by manipulating the
Bohr transition frequency (rather than the laser carrier frequency through the spectral phase) by using a pair of suitably
delayed far-off-resonant Stark-shifting laser pulses, as in SCRAP \cite{SCRAP}.

\subsection{Comparison with earlier work\label{Sec-earlier}}

CPI with a zero-area pulse has been reported in previous publications \cite{CPI earlier}.
However, the physical mechanisms in these publications is different from the one reported here
 because the CPI in the present paper relies on adiabatic time evolution at all times.
By contrast, in the previous proposals \cite{CPI earlier} the sudden change of the phase of the field was essential,
 which led to sudden evolution, in the form of a delta-function shaped nonadiabatic coupling in the adiabatic basis.
This required the use of pulses with large Rabi frequencies, much larger than here.

It is important to note that the zero-area condition is not required in any of these papers because the CPI mechanisms do not require it.
The completely antisymmetric shape of the Rabi frequency, $\Omega(-t)=-\Omega(t)$, and the ensuing zero pulse area, is the most natural realization of
a function that changes sign at time $t=0$, but it is by no means necessary for the physical mechanisms.

\section{Conclusions\label{Sec-Conclusions}}

We have shown that complete population transfer, by means of adiabatic passage, is possible without any crossing of diabatic energy curves.
It is necessary only that the curves touch, and that there be a sign change in the Rabi frequency at that moment.
Such a scenario could be used to replace the conventional two-state rapid adiabatic passage via curve crossing for various implementations.

\acknowledgments
This work has been supported by the European Commission projects EMALI and FASTQUAST,
 and the Bulgarian NSF grants D002-90/08, IRC-CoSiM and Sofia University grant 020/2009.

\subsection*{References}

\end{document}